\documentclass[3p,times,procedia]{elsarticle}
\usepackage{nupha_ecrc}

\volume{00}

\firstpage{1}

\journalname{Nuclear Physics A}

\runauth{}

\jid{nupha}

\jnltitlelogo{Nuclear Physics A}

\usepackage{amssymb}

\usepackage[figuresright]{rotating}

\newcommand{\agev}    {\mbox{~$A$GeV}}               

\newcommand{\mevc}    {\mbox{MeV$/c$}}

\newcommand{\rb}[1]   {\mbox{\textrm{\scriptsize #1}}}

\newcommand{\hefour}  {\ensuremath{{}^{4}\textrm{He}}}

\newcommand{\pt}      {\ensuremath{p_{\rb{t}}}}

\newcommand{\dedx}    {\ensuremath{\textrm{d}E/\textrm{d}x}}

\newcommand{\ycm}     {\ensuremath{y_{\rb{cm}}}}

\begin{document}

\begin{frontmatter}



\dochead{XXVIIth International Conference on Ultrarelativistic Nucleus-Nucleus Collisions\\ (Quark Matter 2018)}

\title{Collective flow and correlations measurements\\ with HADES in Au+Au collisions at 1.23 AGeV}


\author{Behruz Kardan for the HADES Collaboration}
\address{Institut f\"ur Kernphysik, Goethe-Universit\"{a}t, Max-von-Laue-Str. 1, 60438 Frankfurt am Main, Germany}
\ead{bkardan@ikf.uni-frankfurt.de}

\begin{abstract}
The HADES experiment provides a large acceptance combined with a high mass resolution and therefore makes it possible to study dielectron and hadron production in heavy-ion collisions with unprecedented precision. With the high statistics of seven billion Au+Au collisions at 1.23~$A$GeV recorded in 2012 the investigation of collective effects and particle correlations is possible with unprecedented accuracy. We present multi-differential data on directed ($v_1$) and elliptic ($v_2$) flow, and the first measurement of triangular flow ($v_3$), of protons and deuterons.
\end{abstract}

\begin{keyword}

heavy ion collisions \sep light nuclei \sep directed flow \sep elliptic flow \sep triangular flow \sep anisotropic azimuthal correlation \sep nucleon coalescence


\end{keyword}

\end{frontmatter}



\section{Introduction}
Collective flow phenomena are sensitive probes of the general properties of extreme QCD matter~\cite{Ritter:2014uca}, such as its shear viscosity~\cite{Barker:2016hqv}. To achieve a good understanding of these phenomena, flow observables are measured and compared with model calculations to constrain the nuclear Equation-of-State~(EoS)~\cite{Danielewicz:1999zn, Danielewicz:2002pu, Fevre:2015fza,Wang:2018hsw}. The understanding of the EoS of dense matter is of great importance for the investigation of supernovae and compact stars~\cite{Oertel:2016bki}.
In high-energy collisions of nuclei a highly excited nuclear medium is created and its collective expansion produces a correlated emission of particles. In perfectly central collisions the expansion should be isotropic, leading to \emph{radial flow}, which is observable in the transverse-mass spectra of the produced particles. Less central collisions are characterized by an overlap region which is more anisotropic, which results in an azimuthal anisotropy of the momentum space distribution of identified particles. It is common to analyze this by a Fourier decomposition yielding the flow coefficients $v_{1}$, $v_{2}$, $v_{3}$ and higher. Due to their correlation to the collision geometry, the directed $v_{1}$ and elliptic $v_{2}$ are linked to the event plane, which itself is observable via the spectators.
%
\begin{figure}[t]
\begin{center}
\includegraphics[width=0.41\linewidth]{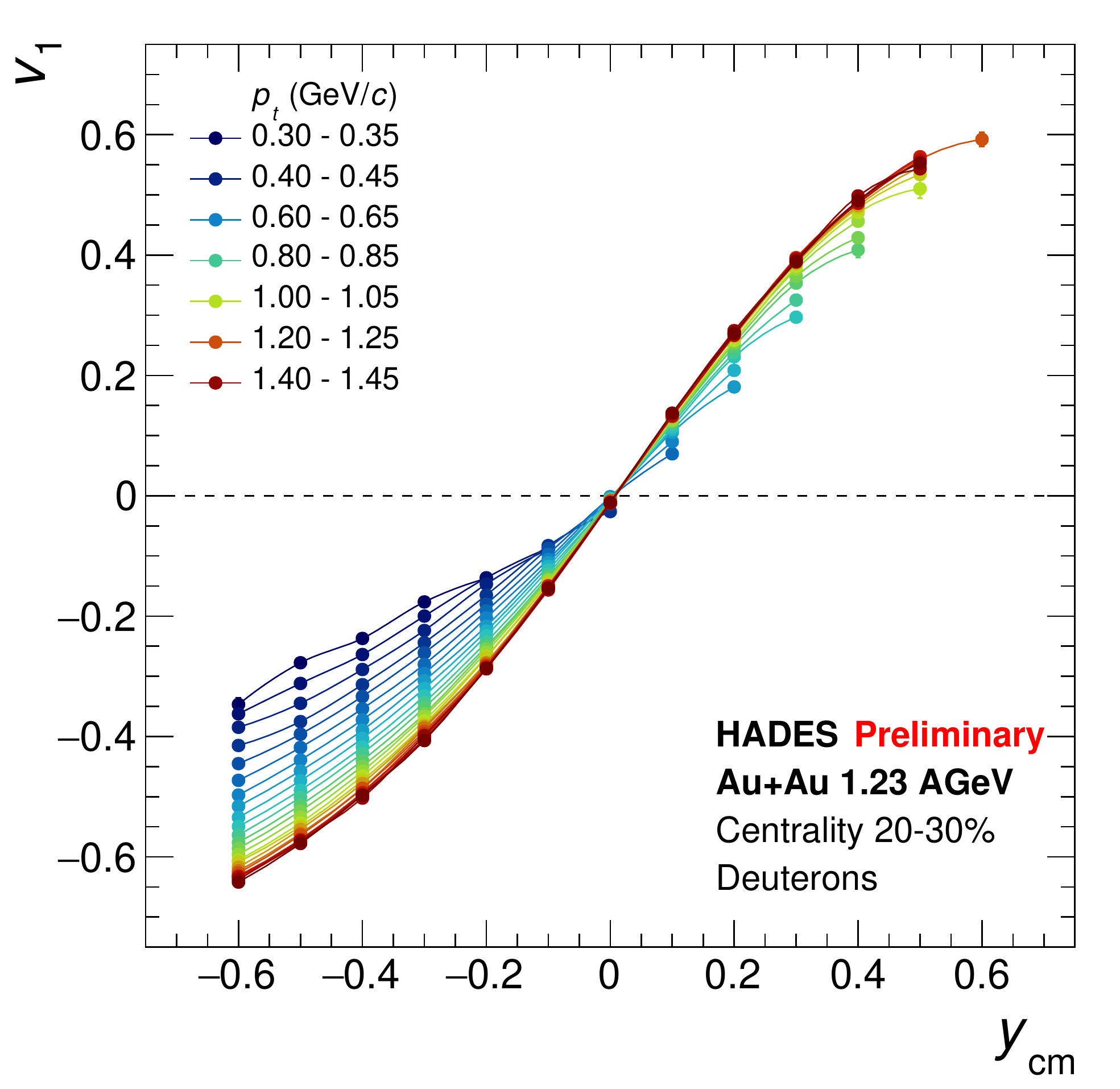}
\hspace{1.5pc}%
\includegraphics[width=0.41\linewidth]{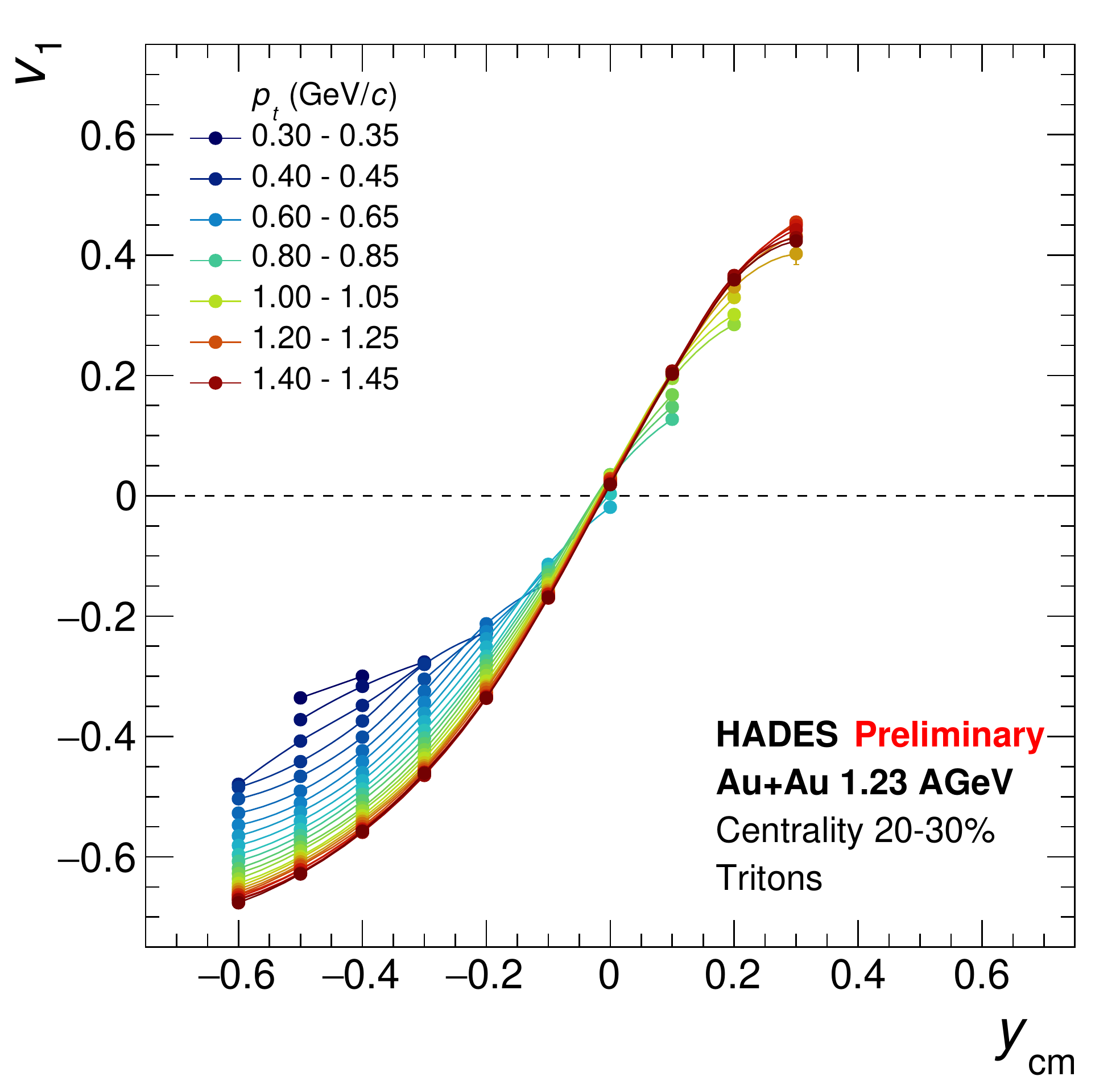}
\includegraphics[width=0.41\linewidth]{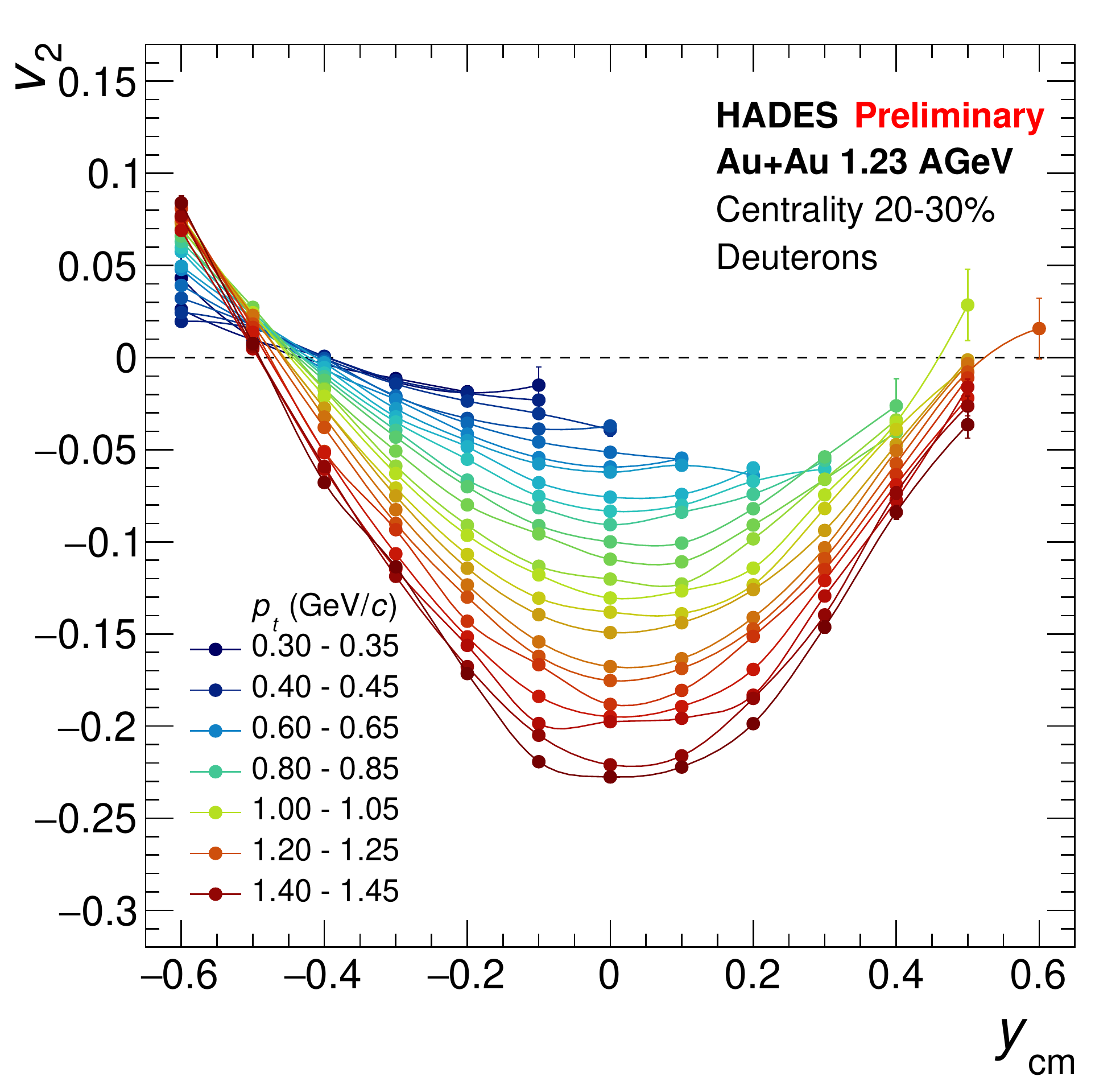}
\hspace{1.5pc}%
\includegraphics[width=0.41\linewidth]{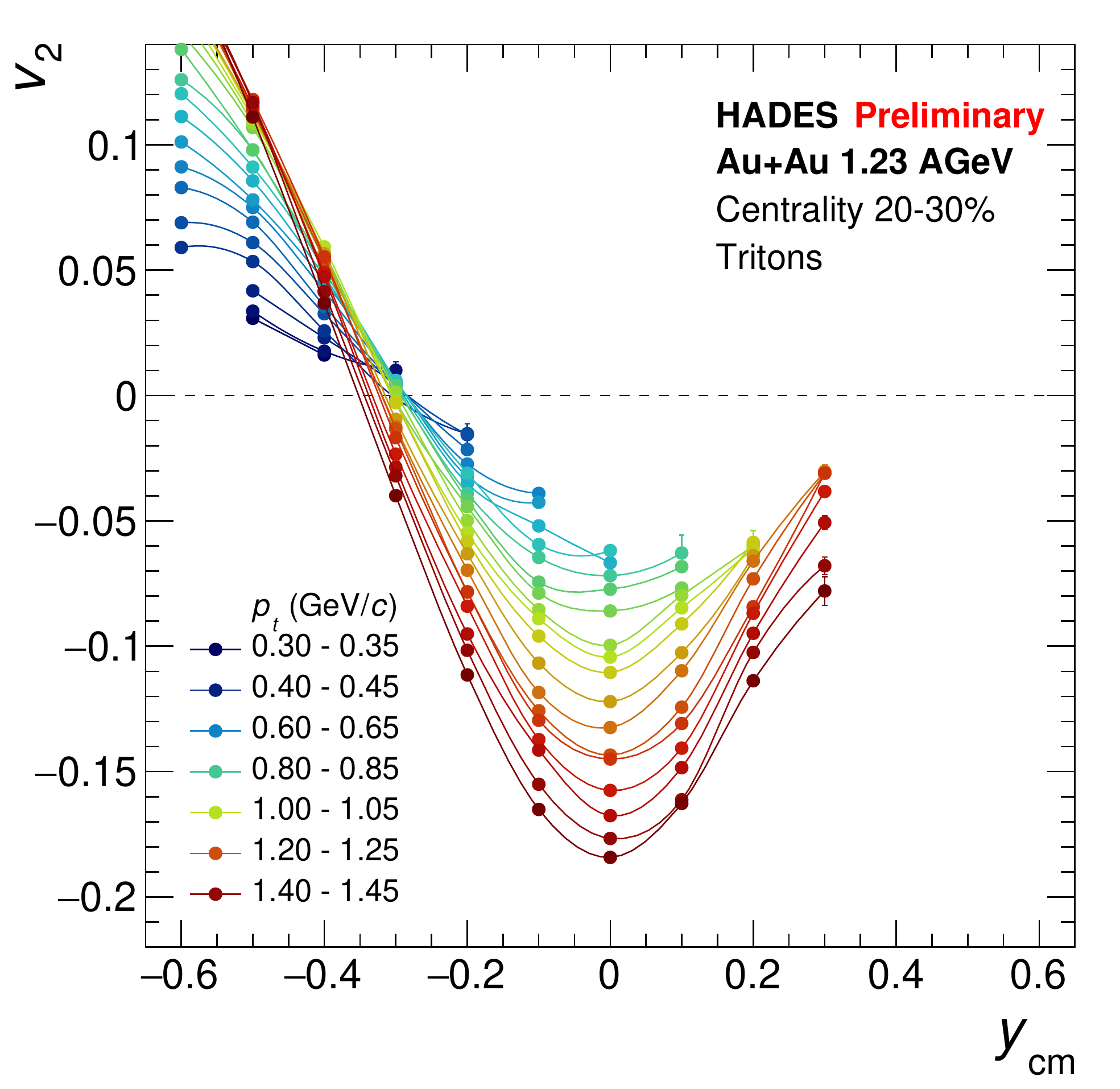}
\end{center}
\caption{Upper panels: directed flow ($v_1$) of (left) deuterons and (right) tritons in semi-central ($20 - 30$~\%) Au+Au collisions at 1.23~$A$GeV as a
  function of the centre-of-mass rapidity \ycm\ in transverse momentum
  intervals of 50~\mevc\ width (lines are to guide the eye).
  In the lower panels the same is shown for the elliptic flow ($v_2$) of (left) deuterons and (right) tritons. Only statistical errors are shown.}
\label{fig:v1v2_d_t_2030Cent}
\end{figure}
%
%
\section{Data set and particle reconstruction}
The High Acceptance DiElectron Spectrometer (HADES) is a fixed-target experiment located at the GSI Helmholtzzentrum f\"ur Schwerionenforschung in Darmstadt, Germany~\cite{Agakishiev:2009am}. The spectrometer is subdivided into six identical sectors, axially symmetric around the beam direction. The momentum reconstruction is carried out by a tracking system consisting in total of 24 multi-wire drift-chambers (MDC), where in each sector two layers are placed in front and two behind a toroidal magnetic field of the superconducting magnet coils. The MDCs also provide the energy loss (\dedx) measurements.  The \emph{Time-Of-Flight} (TOF) and the \emph{Resistive Plate Chamber} (RPC) detector, together with the beam detector (diamond counter), provides the time-of-flight measurements and the trigger information. Particle identification is based on a combined measurement of time-of-flight and energy loss. The additional \dedx~measurement is particularly important to suppress the \hefour~contamination in the deuteron sample, as the two cannot be separated by time-of-flight alone due to the same $Z/A$~ratio. Using simulation the contamination was estimated to be below $1\%$.
The \emph{Forward Wall} (FW), a plastic scintillator hodoscope array, is placed at a distance of $7$~m downstream of the target covering forward angles between $0.3^{\circ}$ and $7.3^{\circ}$ to identify charged projectile spectators by the time-of-flight and their $\Delta E$ signal. The FW hits are used to reconstruct the first-order event plane.
The data analyzed in this work are from the Au+Au run performed in 2012~\cite{Lorenz:2017,Kardan:2017knj}, where the Schwerionen-Synchrotron (SIS) delivered $684$ hours of Au$^{69+}$ ions beam to the HADES experiment~\cite{Bayer:52152}. A fraction of the total recorded events was triggered by selecting events with a charged hit multiplicity in the TOF detector $N_{ch} > 20$, corresponding to $5.85 \times 10^{9}$ events before off-line event selection. According to detailed comparison of the charged track and hit multiplicity distribution with a Glauber Model simulation, this central trigger selects about $43\%$ of the total hadronic cross section of $6.83\pm0.43$~barn, corresponding to a maximum impact parameter of $b_{max}=10$~fm \cite{Centrality}.
\section{Directed and elliptic flow of light nuclei}
\begin{figure}[ht]
\begin{center}
\includegraphics[width=0.41\linewidth]{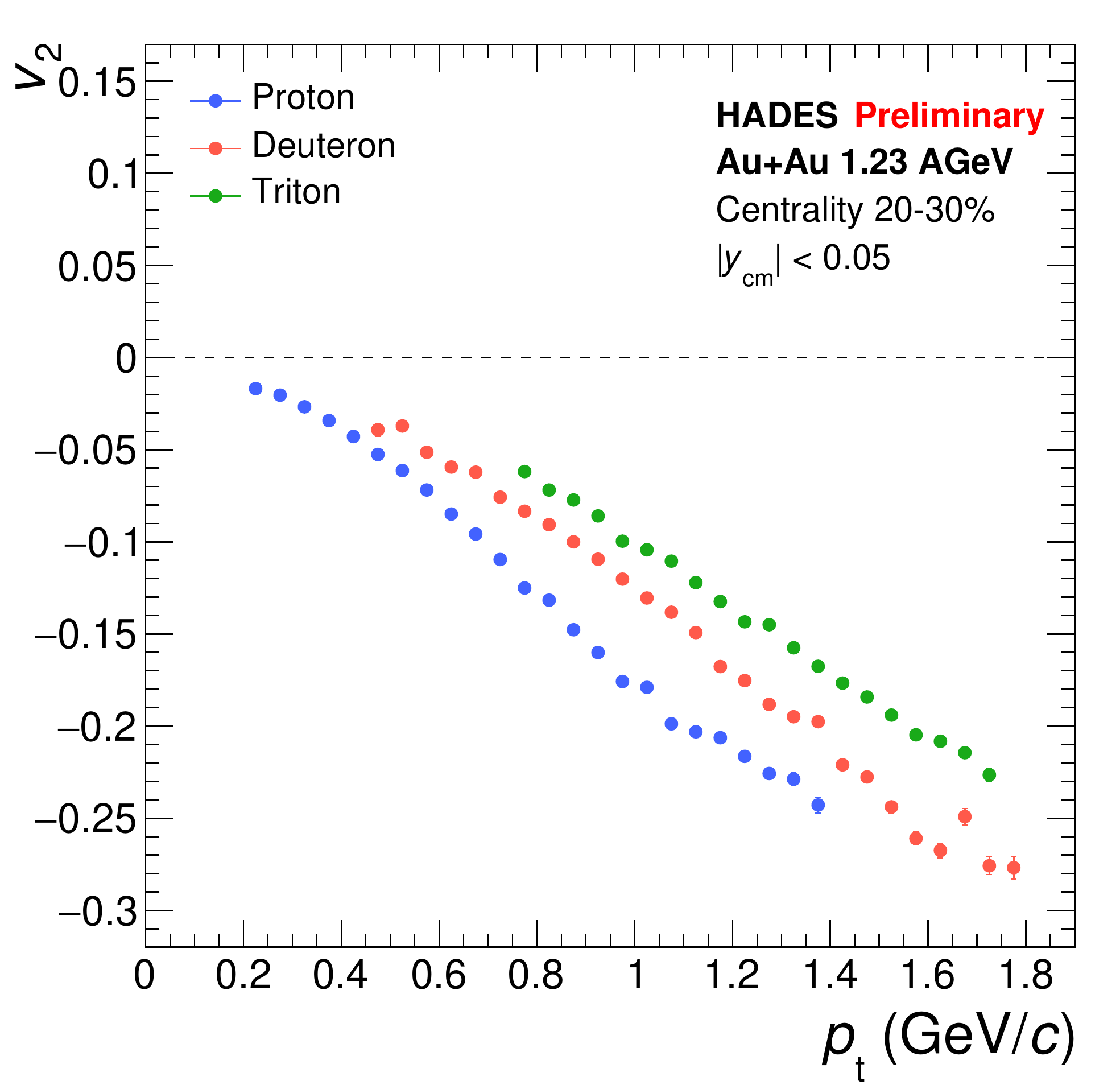}
\hspace{1.5pc}%
\includegraphics[width=0.41\linewidth]{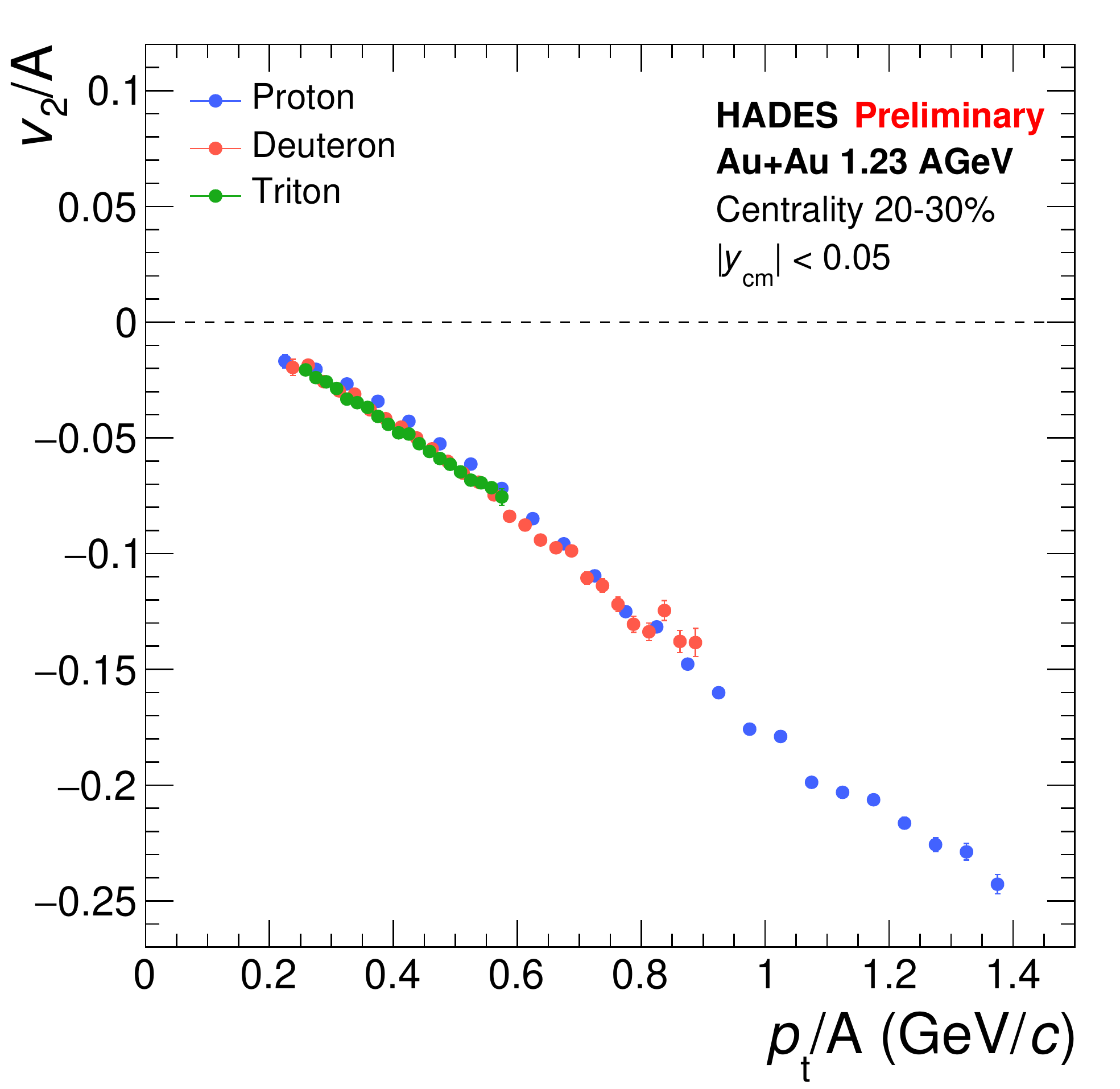}
\end{center}
\caption{Left panel: elliptic flow ($v_2$) of protons, deuterons and tritons as a function of \pt\ around mid-rapidity in semi-central ($20 - 30$~\%) Au+Au collisions at 1.23\agev.  Right panel: The same measurement after scaling $v_2$\ and \pt\ by the mass number $A$ of the nuclei. Only statistical errors are shown here.}
\label{fig:v2A_p_d_t_20-30}
\end{figure}

The directed and elliptic flow ($v_{1}$ and $v_{2}$) of protons (already presented in~\cite{Kardan:2017knj}) and of light nuclei has been extracted over a large region of phase space using the standard event plane method~\cite{Poskanzer:1998yz}. The data have been corrected for the event plane resolution \cite{Ollitrault:1997vz}. In addition, a correction for efficiency losses due to the detector occupancy has been applied track-by-track, as a function of the polar angle, the relative angle to the event plane and the track multiplicity. 
Figure~\ref{fig:v1v2_d_t_2030Cent} shows the results for $v_1$\ and $v_2$\ of deuterons and tritons. A good forward-backward symmetry with respect to mid-rapidity is seen, as expected due to the symmetry of the collision system. Remaining discrepancies are well within the systematic uncertainties, including the systematics due to track reconstruction, particle identification and the non-uniformity of the detector acceptance, which will be presented in detail in an upcoming publication.
Figure~\ref{fig:v2A_p_d_t_20-30} shows a comparison of $v_{2}$ at mid-rapidity for protons, deuterons and tritons. A remarkable scaling of the flow coefficients $v_2$\ is observed after scaling $v_2$\ and \pt\ by the mass number $A$, consistent with naive expectations from nucleon coalescence.
%
%
%
%
\begin{figure}[t]
\begin{center}
\includegraphics[width=0.41\linewidth]{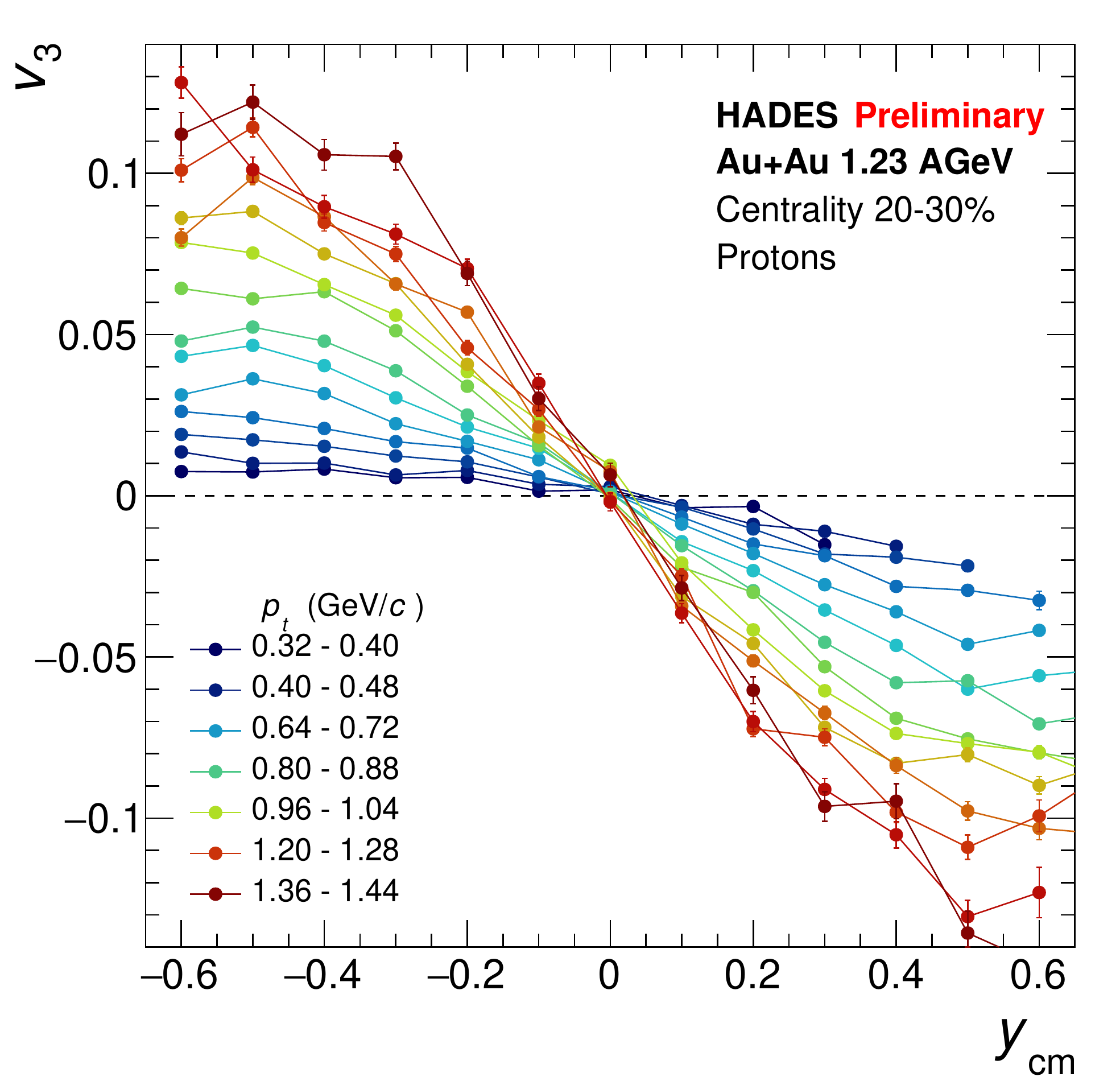}
\hspace{1.5pc}%
\includegraphics[width=0.41\linewidth]{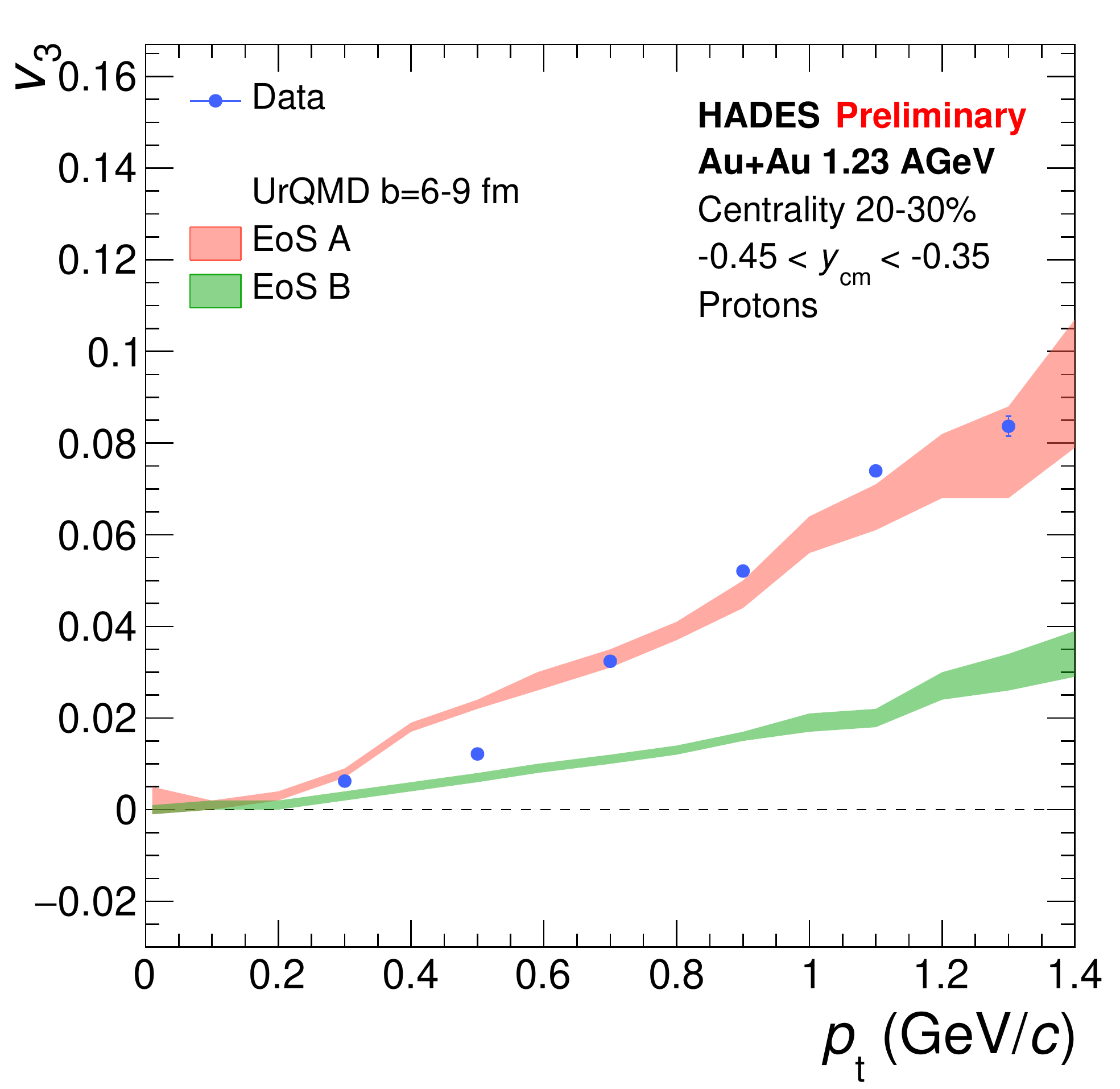}
\includegraphics[width=0.41\linewidth]{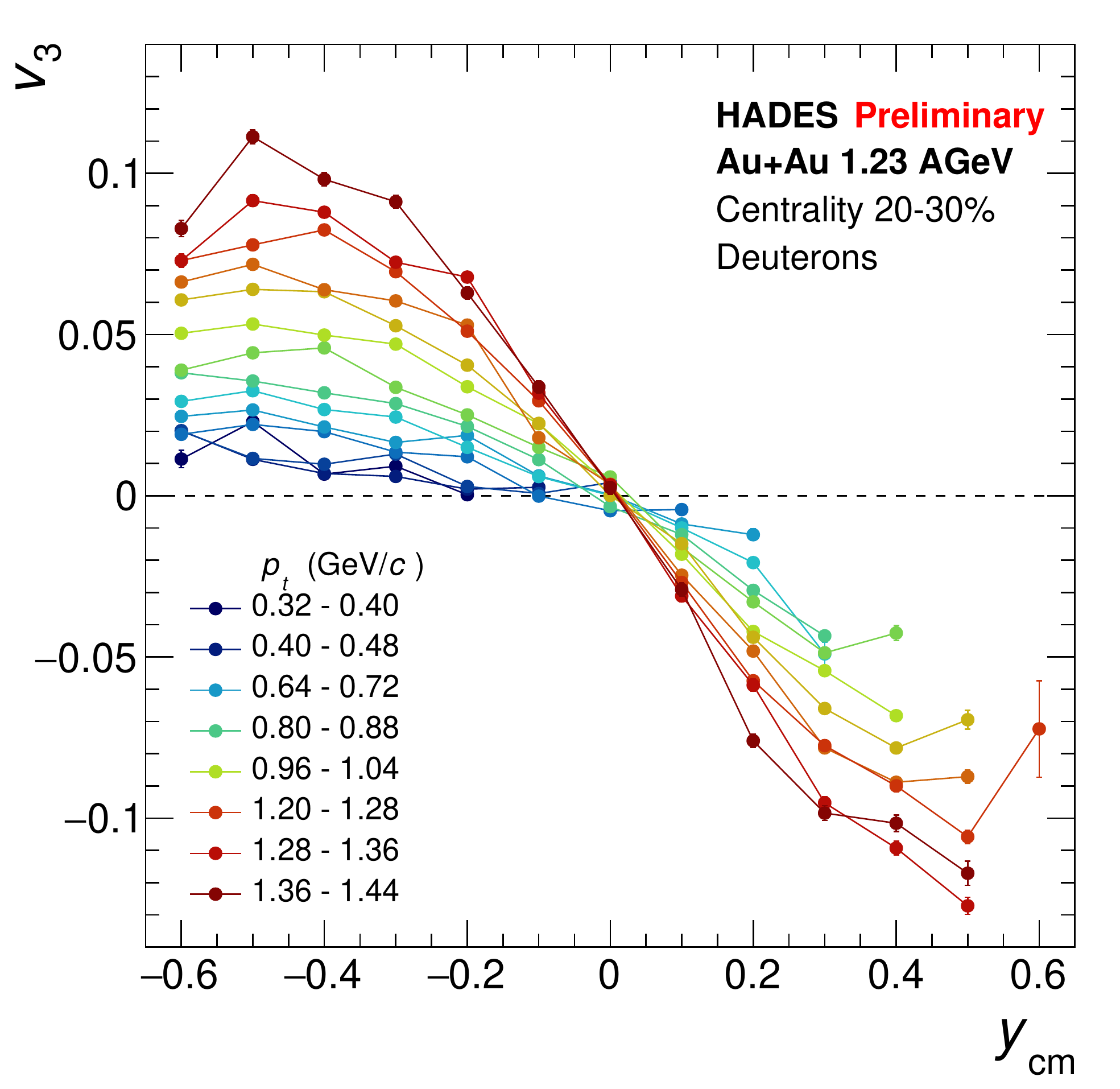}
\hspace{1.5pc}%
\includegraphics[width=0.41\linewidth]{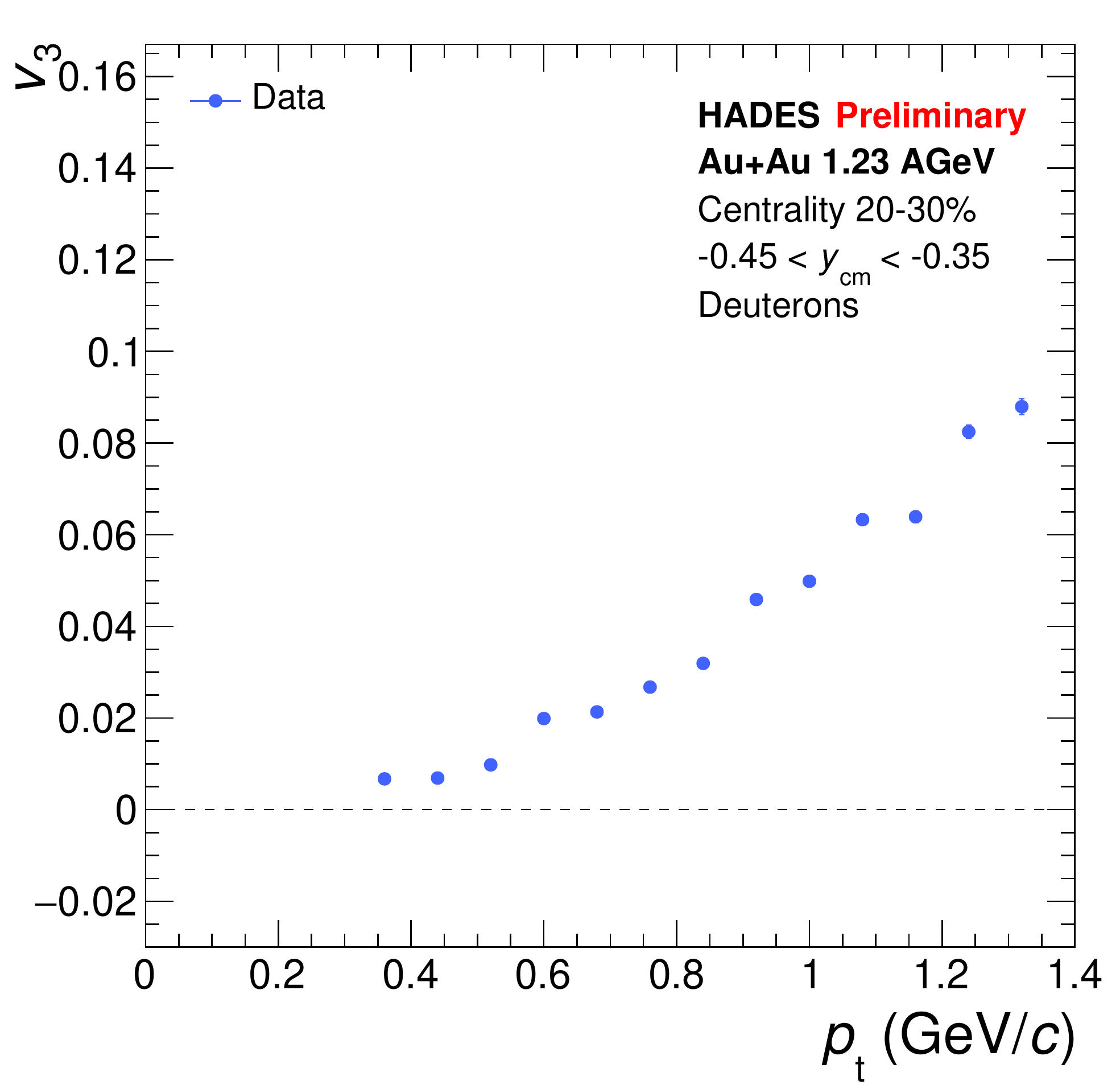}
\end{center}
\caption{Upper left panel: triangular flow ($v_3$) of protons in
  semi-central ($20 - 30$~\%) Au+Au collisions at 1.23~$A$GeV as a
  function of the centre-of-mass rapidity \ycm\ in transverse momentum
  intervals of 80~\mevc\ width (lines are to guide the eye).
  Upper right panel: proton $v_3$\ as a function of \pt\ in the
  rapidity interval $-0.45 < y_{cm} < -0.35$ together with UrQMD3.4 simulations with two parametrization of EoS~\cite{Hillmann:2018nmd}.
     In the lower two panels the same is shown for the triangular flow ($v_3$) of deuterons.
     Only statistical errors are shown here.}
\label{fig:v3_p_d_20-30}
\end{figure}
%
%
\section{Higher flow coefficients: $v_3$}
A first measurement at SIS energies of the triangular flow coefficients relative to the first order event plane ($v_3$) of protons and deuterons is shown in Fig.~\ref{fig:v3_p_d_20-30}. The observation of triangular flow at high energies is interpreted as the outcome of initial state fluctuations~\cite{Alver:2010gr}, which, however, are not correlated to the event plane. In contrast to this, the measurement shown here is done w.r.t. the first-order event plane and thus $v_3$\ should have a different origin.
In UrQMD3.4~\cite{Bleicher:1999xi} simulations (see Fig.~\ref{fig:v3_p_d_20-30}) a strong sensitivity to the EoS is observed which in this case is expected to be even more sensitive to the strength of the used potential than $v_1$\ and $v_2$~\cite{Hillmann:2018nmd}.
\\
\\
Work supported by BMBF (05P15RFFCA), GSI, HGS-HIRe, H-QM and HICforFAIR.




\bibliographystyle{elsarticle-num}



\end{document}